\documentclass[dvips]{acta}
\usepackage{supertabular,lscape,epsfig}
\usepackage{amssymb}
\usepackage{amsmath}
\DeclareSymbolFont{ppa}{OT1}{ppl}{m}{it}
\DeclareMathSymbol{\vv}{\mathalpha}{ppa}{'166}

\newfont{\hb}{rphvb at 10pt}
\newfont{\hbo}{rphvbo at 10pt}
\newfont{\bitt}{rptmbi at 12pt}
\newfont{\bits}{rptmbi at 11pt}

\SetPages{1}{1}

\SetVol{64}{2014}

\begin{document}

\newcommand{\TabCapp}[2]{\begin{center}\parbox[t]{#1}{\centerline{
  \small {\spaceskip 2pt plus 1pt minus 1pt T a b l e}
  \refstepcounter{table}\thetable}
  \vskip2mm
  \centerline{\footnotesize #2}}
  \vskip3mm
\end{center}}

\newcommand{\TTabCap}[3]{\begin{center}\parbox[t]{#1}{\centerline{
  \small {\spaceskip 2pt plus 1pt minus 1pt T a b l e}
  \refstepcounter{table}\thetable}
  \vskip2mm
  \centerline{\footnotesize #2}
  \centerline{\footnotesize #3}}
  \vskip1mm
\end{center}}

\newcommand{\MakeTableSepp}[4]{\begin{table}[p]\TabCapp{#2}{#3}
  \begin{center} \TableFont \begin{tabular}{#1} #4
  \end{tabular}\end{center}\end{table}}

\newcommand{\MakeTableee}[4]{\begin{table}[htb]\TabCapp{#2}{#3}
  \begin{center} \TableFont \begin{tabular}{#1} #4
  \end{tabular}\end{center}\end{table}}

\newcommand{\MakeTablee}[5]{\begin{table}[htb]\TTabCap{#2}{#3}{#4}
  \begin{center} \TableFont \begin{tabular}{#1} #5
  \end{tabular}\end{center}\end{table}}

\newfont{\bb}{ptmbi8t at 12pt}
\newfont{\bbb}{cmbxti10}
\newfont{\bbbb}{cmbxti10 at 9pt}
\newcommand{\uprule}{\rule{0pt}{2.5ex}}
\newcommand{\douprule}{\rule[-2ex]{0pt}{4.5ex}}
\newcommand{\dorule}{\rule[-2ex]{0pt}{2ex}}
\def\thefootnote{\fnsymbol{footnote}}
\begin{Titlepage}
\Title{Ultra-Short-Period Binary Systems\\
in the OGLE Fields Toward the Galactic Bulge\footnote{Based on
observations obtained with the 1.3-m Warsaw telescope at the Las
Campanas Observatory of the Carnegie Institution for Science.}}
\Author{I.~~S~o~s~z~y~ñ~s~k~i$^1$,~~
K.~~S~t~ê~p~i~e~ñ$^1$,~~
B.~~P~i~l~e~c~k~i$^{1,2}$,~~
P.~~M~r~ó~z$^1$,~~
A.~~U~d~a~l~s~k~i$^1$,\\
M.\,K.~~S~z~y~m~a~ñ~s~k~i$^1$,~~
G.~~P~i~e~t~r~z~y~ñ~s~k~i$^{1,2}$,~~
\L.~~W~y~r~z~y~k~o~w~s~k~i$^{1,3}$,\\
K.~~U~l~a~c~z~y~k$^1$,~~
R.~~P~o~l~e~s~k~i$^{1,4}$,~~
S.~~K~o~z~³~o~w~s~k~i$^1$,~~
P.~~P~i~e~t~r~u~k~o~w~i~c~z$^1$,\\
J.~~S~k~o~w~r~o~n$^1$~~
and~~M.~~P~a~w~l~a~k$^1$}
{$^1$Warsaw University Observatory, Al.~Ujazdowskie~4, 00-478~Warszawa, Poland\\
e-mail: soszynsk@astrouw.edu.pl\\
$^2$ Universidad de Concepción, Departamento de Astronomia, Casilla 160--C, Concepción, Chile\\
$^3$ Institute of Astronomy, University of Cambridge, Madingley Road, Cambridge CB3 0HA, UK\\
$^4$ Department of Astronomy, Ohio State University, 140 W. 18th Ave., Columbus, OH 43210, USA}
\Received{~}
\end{Titlepage}
\Abstract{We present a sample of 242 ultra-short-period
($P_{\mathrm{orb}}<0.22$~d) eclipsing and ellipsoidal binary stars
identified in the OGLE fields toward the Galactic bulge. Based on the
light curve morphology, we divide the sample into candidates for
contact binaries and non-contact binaries. In the latter group we
distinguish binary systems consisting of a cool main-sequence star and
a B-type subdwarf (HW~Vir stars) and candidates for cataclysmic
variables, including five eclipsing dwarf novae. One of the detected
eclipsing binary systems -- OGLE-BLG-ECL-000066 -- with the orbital
period below 0.1~d, likely consists of M dwarfs in a nearly contact
configuration. If confirmed, this would be the shortest-period M-dwarf
binary system currently known. We discuss possible evolutionary
mechanisms that could lead to the orbital period below 0.1~d in an
M-dwarf binary.}
{binaries: eclipsing -- stars: low-mass -- novae, cataclysmic
variables -- stars: individual: OGLE-BLG-ECL-000066}

\Section{Introduction}

In recent years the list of known eclipsing binary systems with very short
periods has been substantially extended. It is known for more than two
decades (Rucinski 1992) that contact binaries have a sharp cut-off at
$\sim0.22$~d in the period distribution, and very few systems have periods
significantly below this limit. For years, the shortest-period known binary
system with M-dwarf components was OGLE BW3 V38 ($P=0.1984$~d) discovered
by Udalski \etal (1995) and studied in detail by Maceroni and Rucinski
(1997). This record was beaten by Norton \etal (2007), who reported the
discovery of a binary system with main sequence components on a
$P=0.1926$-d orbit. Then, rich samples of ultra-short period binaries were
published by Norton \etal (2011), Nefs \etal (2012), Lohr \etal (2013), and
recently by Drake \etal (2014), who identified as many as 367 binary
systems with periods below 0.22~days using photometric databases collected
by the Catalina Surveys. Until now, the shortest known period of the M
dwarf binary systems was $P=0.1122$~d (Nefs \etal 2012).

Two main hypotheses have recently been proposed to explain the short-period
cut-off in contact eclipsing binaries. Stêpieñ (2006b) argued that the
short-period limit of contact binaries may be caused by the fact that the
initially detached systems with low-mass components do not have time to
reach the Roche lobe overflow within the age of the Universe. An existence
of a lower limit about 2~d for the initial orbital period is essential for
this explanation. It results from a binary star formation mechanism in
which a binary is formed due to a fragmentation of a protostellar cloud
(Bonnell 1994, Machida \etal 2008, Kratter \etal 2010). So, a new born cool
binary consists of two T~Tau stars and its orbit must be wide enough to
accommodate components with radii of a few solar radii. A similar low limit
for a period of young binaries is expected when the mechanism called Kozai
cycles with a tidal friction (KCTF) operates on a binary with an initially
longer period (Fabrycky and Tremaine 2007, Perets and Fabrycky 2009, Naoz
and Fabrycky 2014). For components with masses close to 0.2 $M_{\odot}$,
the low limit for the initial period is somewhat lower -- around 1.5~d
(Nefs \etal 2012). This does not mean, of course, that shorter initial
periods are completely forbidden. The KCTF mechanism can produce shorter
periods under exceptional circumstances and also a close approach with
another body can ``harden'' a binary, shortening significantly its period
(e.g. Hypki and Giersz 2013). Nevertheless, young binaries with very short
periods are expected to be rare.

On the other hand, Jiang \etal (2012) noted that existence of detached and
semi-detached binary systems with periods below the short-period limit are
in conflict with the suggestion of Stêpieñ (2006b). The existence of
systems with such short periods indicates that some of the binaries may
have short periods at their birth or that they may experience a much higher
angular momentum loss (AML) than assumed by Stêpieñ (2006b). According to
Jiang \etal (2012), the short-period limit of contact binaries results from
a limit for the mass of the primary component. For initial primary mass
lower than 0.63~$M_\odot$, the mass transfer that starts, when the primary
reaches its Roche lobe, is dynamically unstable, which quickly leads to the
common envelope binary and the coalescence of both components. Only
binaries with the primary mass higher than 0.63~$M_\odot$ may form
long-lived W~UMa stars and their orbital periods are longer than about
0.22~d. This explanation may be invoked if, indeed, a substantial fraction
of young binaries has very short periods of 1~d or less.

The mystery of the short-period cut-off could be solved by detailed
examination of the ultra-short-period binaries. The relative numbers of
contact versus detached binaries and their period distribution may be an
important constraint on models of the formation and migration history in
low-mass binary systems. Analysis of the orbital period changes could
answer the question of whether the AML in low-mass close binaries is so
fast that such systems quickly merge and form single stars (Jiang \etal
2012) or, on the contrary, the AML is so feeble that any M-dwarf binary had
no time to form contact system (Stêpieñ 2006b). For binaries with orbits
tidally tightened by the KCTF mechanism the third, circumbinary, objects
should also be detected through the timing analysis.

Therefore, it is essential to find as many as possible ultra-short-period
M-dwarf binaries to test the predictions of angular momentum loss
theories. In the present study we publish the list of 242 binary systems
with orbital periods below 0.22~d detected toward the Optical Gravitational
Lensing Experiment (OGLE) fields in the Galactic bulge. Our sample consists
not only of the M-dwarf binaries, but also of the main-sequence -- hot
sub-dwarf systems and of cataclysmic variables. One of our objects --
OGLE-BLG-ECL-000066 -- is probably the shortest-period binary system
consisting of two M-dwarfs.

\Section{Observations and Data Reduction}
OGLE is a long-term photometric survey for sky variability operating on the
1.3-m Warsaw Telescope at Las Campanas Observatory, Chile. The observatory
is operated by the Carnegie Institution for Science. In this investigation,
we used photometric data collected during the third and the fourth phases
of the OGLE survey (OGLE-III and OGLE-IV) in the years 2001--2013.
Currently, the telescope is equipped with a 32-chip 256 megapixel mosaic
camera with 1.4 square degrees field of view. The OGLE survey uses two
photometric filters: about 90\% of the observations are obtained in the
Cousins {\it I} band, the remaining images were made with the Johnson {\it
V} band. The number of observations substantially varies from field to
field: from about 100 to more than 8000 points per light curve.

The short-period eclipsing binary candidates were detected in the area
of 182 square degrees covering central regions of the Milky Way. The
{\it I}-band magnitudes in the OGLE-IV database range from about
13~mag to 20.5~mag. Data reduction of the OGLE images is performed
using the Difference Image Analysis pipeline (Alard and Lupton 1998,
Wo¼niak 2000). Detailed descriptions of the instrumentation,
photometric reductions and astrometric calibrations of the OGLE data
are available in Udalski (2003a) and Udalski \etal (2008).

\Section{Selection and Classification of Ultra-Short-Period Binary Systems}

The period analysis for nearly 400 million {\it I}-band light curves
collected by OGLE in the Galactic bulge was done using {\sc Fnpeaks}
code\footnote{see
http://helas.astro.uni.wroc.pl/deliverables.php?lang=en\&active=fnpeaks}
written by Z.~Ko³acz\-kowski. We searched a frequency space from 0 to
24~d$^{-1}$ with a spacing of 0.00005~d$^{-1}$. The light curves with
periods below 0.22~d and with the highest signal-to-noise ratio
were subjected to visual inspection. Identified variable stars were
divided into three groups: eclipsing/ellipsoidal binaries, pulsating
stars (in most cases $\delta$~Sct stars) and other variables. The
latter group contains mostly variables of undetermined types. Some of
them may be binary systems or pulsating stars, but we are not able to
confidently classify them based solely on the OGLE light curves.

In this paper, we present 242 objects gathered in the first group --
containing rather certain cases of short-period eclipsing or
ellipsoidal variables. In order to classify our binary systems into
different types, each {\it I}-band light curve was analyzed with the
aid of the Wilson-Devinney light curve modeling technique (Wilson and
Devinney 1971, hereafter WD code). We have used the procedure
developed by Pilecki (2010), see also Pilecki and Stêpieñ (2012). Our
method is based on a mixture of standard Monte Carlo approach and the
Markov Chain Monte Carlo (MCMC) method. In this approach, model light
curves were created using the WD code for a given random parameter
set. Then they were compared with the observational data and the
reduced $\chi^2$ values were calculated to measure the goodness of the
model. The final model was the one with the lowest $\chi^2$ value.

For the purpose of this work the method was simplified, as we did not
have radial velocities measured for the stars. In the analysis we
included the following parameters: the modified surface potentials
$\Omega_1$ and $\Omega_2$, the orbital inclination $i$, the
temperature ratio $T_2/T_1$ (temperature of the hotter component was
fixed), and the phase shift. The period and the reference time of the
primary minimum was taken from the Fourier analysis of the light
curves, but the phase shift took care of any error in the
latter. Also, for each star two mass ratios were tested, one
representing the components of equal masses ($q=1.0$) and the other
representing components of unequal masses ($q \sim 0.35$), but they
were fixed during the analysis. For individual stars other values of
mass ratios were also tested, if needed.

To facilitate the automated search for the best model we divided the
sample into two groups -- one with narrow eclipses (meaning higher
component separation) and no distinct proximity effects visible --
detached system candidates, and the other with wider eclipses and a
continuous brightness change (with proximity effects) -- compact
system candidates. For both groups we used a different set of initial
parameters and parameter ranges in order to minimize necessary
calculations to find the global minimum. For the detached system
group, we started from a higher component separation and higher
inclination angles and for the compact system group from a lower
component separation and in a wider range of inclination angles. Also,
for some detached systems with extreme ratios of eclipse depths, we
changed the initial temperature ratio of the model, and for stars with
distinct reflection effect (see Section 5.3) the temperature of the
hotter component was set to 30,000~K. In other cases the temperature
was set to 5000~K (compact systems) and 6000~K (detached systems), as
it has only an indirect influence by the limb darkening coefficients
and does not affect much the solution.

The following steps were practically the same for all the stars. Using
a slightly modified standard Monte Carlo approach we were looking for
the best models in a wide range of parameter space. To avoid situation
in which the global minimum is outside the parameter range instead of
using the uniform distribution, we were taking values from the normal
distribution minimizing this risk. Once about a thousand models are
calculated, the range and the central point of the distribution are
automatically changed to better cover the parameter ranges for which
the best models are obtained. Narrowing the parameter ranges (changing the
sigma and the central value of the normal distribution) we were able
to find a global minimum in the $\chi^2$ plane. Between some steps the
MCMC method was also applied to minimize the risk of being stuck at
some local minimum. In this method there is no fixed range of
parameters and no central point of distribution is used, so after a
sufficient steps it may virtually cover the whole range of parameters
value. There is a restriction though that makes this method avoid
parameter sets that give highly improbable models.

Using this procedure, calculating a few thousand models was in general
sufficient to find a global minimum with enough accuracy and even
estimate quite reliably the errors of the parameters (taking into
account all the simplifications). For each system the reduced $\chi^2$
values for different mass ratios were compared and the model with
the lowest value was selected as the best one. Calculating models for more
values of mass ratios could be done, but the two tested values were
sufficient to fulfill the main goal of this analysis -- to obtain the
approximate parameters and to tell if the stars are in contact or
not. Also one must have in mind that the photometric mass ratios are
just a rough approximation anyway and that they can be trusted only in
some specific cases (contact binaries with total eclipses), since in
general one needs a spectroscopic mass ratio. The configuration of the
system was determined using the Roche lobe filling factor of the
components. Here we measure this factor as a ratio between the polar
radius of the star and the polar radius of the corresponding Roche
lobe. If these ratios for both components are higher or equal to one,
the system is considered as a contact one. These objects display light
curves that are characteristic of W~UMa-type stars: similar
depths of minima with continuously changing brightness because of
large tidal distortion of the two components. Other cases --
semi-detached and detached systems -- were grouped into one class --
non-contact binaries. Among these objects we could distinguish
interesting sub-groups: cataclysmic variables and systems containing
cool main sequence stars and hot evolved companions. Also, using the
inclination parameters and the relative radii of the stars calculated
from the potentials, we were able to tell if there are no eclipses at
all, and that the brightness change is only or mostly due to the
ellipsoidal variation.

\begin{figure}[t]
\includegraphics[width=12.5cm]{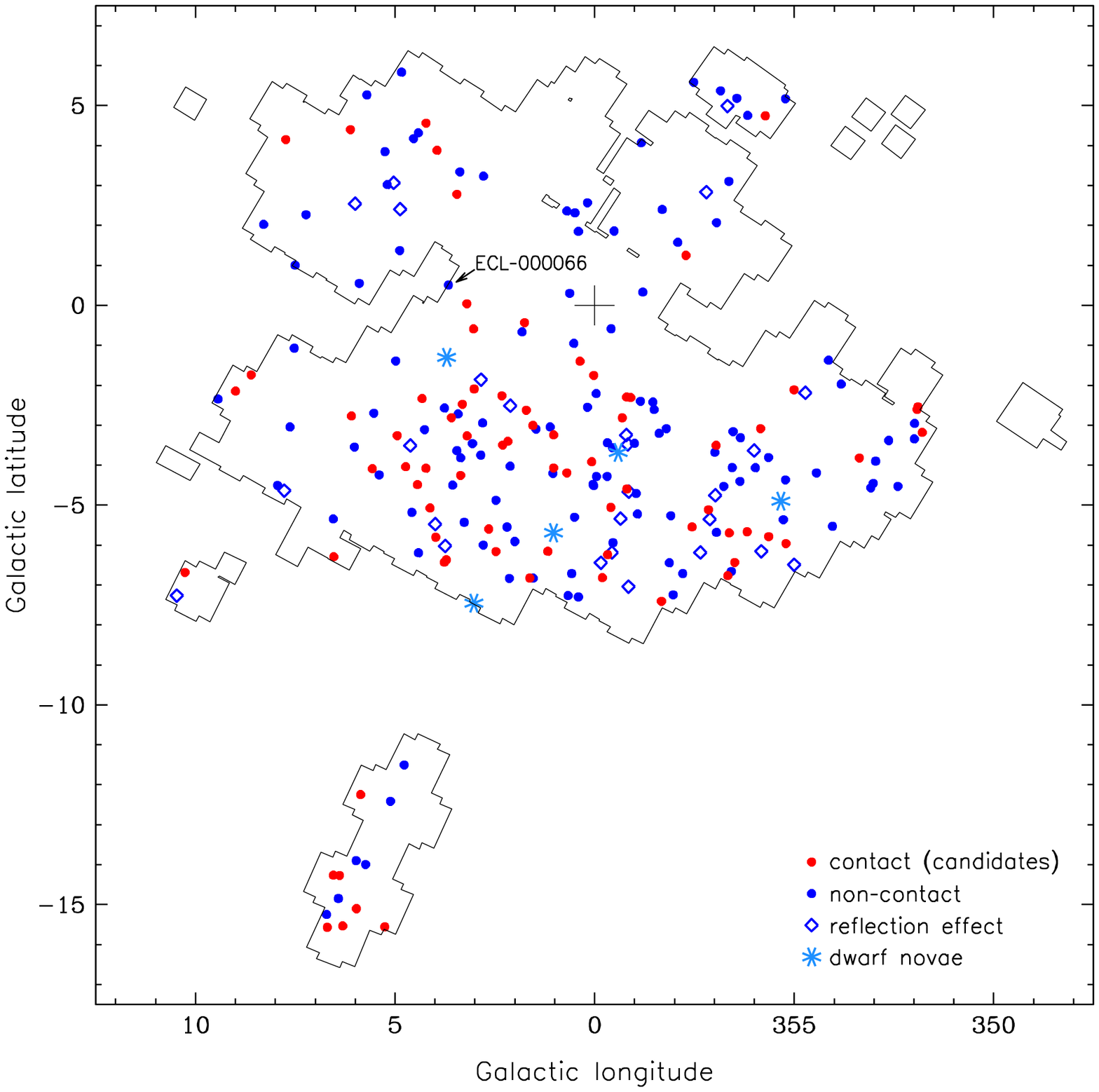}
\FigCap{Spatial distribution of ultra-short-period binary systems in the
OGLE fields toward the Galactic bulge. Red points indicate candidates for
contact binaries, blue points show positions of non-contact (detached and
semi-detached) systems. Stars with strong reflection effect are marked with
empty diamonds, dwarf novae are showed with blue stars. Black
line show contours of the OGLE fields toward the Galactic bulge.}
\end{figure}

\begin{figure}[t]
\includegraphics[width=12.5cm]{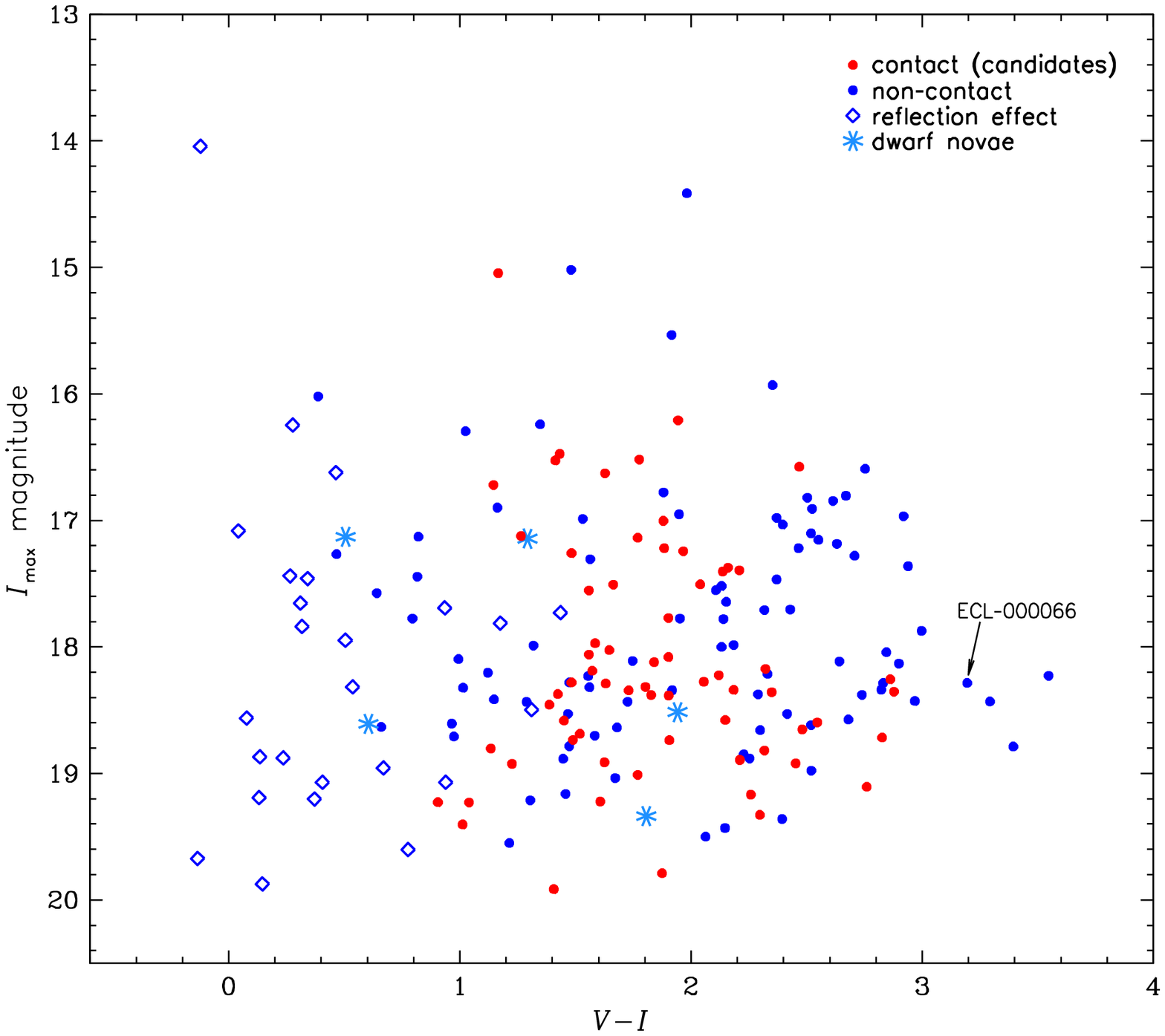}
\FigCap{Color--magnitude diagram for ultra-short-period binary
systems. Different symbols have the same meaning as in Fig.~1.}
\end{figure}

\Section{OGLE Sample of Ultra-Short-Period Binary Systems}

The final sample consists of 242 short-period eclipsing and
ellipsoidal binary systems detected toward the Galactic bulge. 75 of
these objects are candidates for eclipsing contact binaries (W~UMa
stars), 167 stars are probable semi-detached and detached
systems. From the latter group, five stars are dwarf novae, four stars
are known Novae and ten other objects are candidates for cataclysmic
variables. We provide basic observational parameters of our sample,
the OGLE time-series photometry in the {\it I} and {\it V} bands and
finding charts. These data can be downloaded from the FTP anonymous
site:

\begin{center}
{\it ftp://ftp.astrouw.edu.pl/ogle/ogle4/OCVS/blg/short\_period\_ecl/}
\end{center}

Eclipsing and ellipsoidal systems were given designations
OGLE-BLG-ECL-NNNNNN and OGLE-BLG-ELL-NNNNNN, respectively, where
NNNNNN is a six-digit number. This designation scheme will be
continued in the OGLE collection of eclipsing and ellipsoidal
variables, which will be published in the future. Table~1 contains
first 30 lines the file {\sf list.dat} from the FTP site. For each
star we provide its identifiers, J2000 equatorial coordinates, type of
variability (C -- contact binary, NC -- non-contact binary, ELL --
ellipsoidal variable, CV -- cataclysmic variable), {\it I}- and {\it
V}-band magnitudes at maximum light, orbital period, primary and
secondary eclipse depths in the {\it I}-band, and epoch of the primary
eclipse minimum. The orbital periods were refined with the {\sc Tatry}
code (Schwarzenberg-Czerny 1996).

\begin{landscape}
\renewcommand{\TableFont}{\scriptsize}
\MakeTable{l@{\hspace{5pt}}
l@{\hspace{2pt}}
c@{\hspace{3pt}}
c@{\hspace{1pt}}
c@{\hspace{1pt}}
c@{\hspace{3pt}}
c@{\hspace{3pt}}
c@{\hspace{1pt}}
c@{\hspace{-3pt}}
c@{\hspace{-1pt}}
c}{12.5cm}{First 30 lines of the file {\sf list.dat}.}
{\hline
\multicolumn{1}{c}{Identifier} 
&\multicolumn{1}{c}{OGLE} 
&\multicolumn{1}{c}{R.A.} 
&\multicolumn{1}{c}{Dec} 
&\multicolumn{1}{c}{Type} 
&\multicolumn{1}{c}{$I_{\mathrm{max}}$} 
&\multicolumn{1}{c}{$V_{\mathrm{max}}$} 
&\multicolumn{1}{c}{$P_{\mathrm{orb}}$} 
&\multicolumn{1}{c}{$A_1(I)$} 
&\multicolumn{1}{c}{$A_2(I)$} 
&\multicolumn{1}{c}{$T_{\mathrm{min}}$} \\
&\multicolumn{1}{c}{Identifier} 
&\multicolumn{1}{c}{[J2000.0]} 
&\multicolumn{1}{c}{[J2000.0]} & 
&\multicolumn{1}{c}{[mag]} 
&\multicolumn{1}{c}{[mag]} 
&\multicolumn{1}{c}{[d]} 
&\multicolumn{1}{c}{[mag]} 
&\multicolumn{1}{c}{[mag]} 
& HJD-2450000 \\
\hline
OGLE-BLG-ECL-000001 & BLG617.07.79075  & 17:13:40.94 & $-$30:04:17.2 &  NC & 18.714 &   --   & 0.20903930 & 0.916 & 0.538 & 5000.15657 \\
OGLE-BLG-ECL-000002 & BLG617.11.65828  & 17:16:38.36 & $-$29:54:33.6 &   C & 17.244 & 19.210 & 0.21367127 & 0.440 & 0.386 & 5000.04525 \\
OGLE-BLG-ECL-000003 & BLG616.03.6996   & 17:16:51.98 & $-$29:04:29.1 &  NC & 19.057 &   --   & 0.21434213 & 0.840 & 0.625 & 5000.14498 \\
OGLE-BLG-ECL-000004 & BLG616.10.93602  & 17:17:13.44 & $-$28:38:07.1 &  NC & 17.708 & 20.025 & 0.21488204 & 0.970 & 0.771 & 5000.03817 \\
OGLE-BLG-ECL-000005 & BLG617.19.15379  & 17:17:46.19 & $-$29:32:42.2 &  NC & 14.736 &   --   & 0.21549714 & 0.782 & 0.509 & 5000.18320 \\
OGLE-BLG-ECL-000006 & BLG616.26.30381  & 17:18:11.28 & $-$27:58:02.9 &  NC & 19.063 &   --   & 0.21877314 & 0.950 & 0.750 & 5000.10010 \\
OGLE-BLG-ECL-000007 & BLG616.01.26238  & 17:18:11.61 & $-$28:59:37.3 &  NC & 19.601 & 20.377 & 0.09176723 & 0.970 & 0.197 & 5000.03653 \\
OGLE-BLG-ECL-000008 & BLG615.11.36403  & 17:25:16.06 & $-$30:05:26.2 &  NC & 17.519 & 19.651 & 0.21770001 & 0.576 & 0.407 & 5000.16501 \\
OGLE-BLG-ECL-000009 & BLG612.25.57960  & 17:27:10.66 & $-$27:43:57.5 &  NC & 18.035 &   --   & 0.15077164 & 0.342 & 0.159 & 5000.00129 \\
OGLE-BLG-ECL-000010 & BLG613.07.75589  & 17:27:44.28 & $-$29:46:16.1 &  NC & 16.621 & 17.085 & 0.10051416 & 1.271 & 0.165 & 5000.01173 \\
OGLE-BLG-ECL-000011 & BLG662.21.5107   & 17:30:01.74 & $-$30:24:29.8 &  NC & 17.775 & 19.727 & 0.20005222 & 0.526 & 0.472 & 5000.03194 \\
OGLE-BLG-ECL-000012 & BLG613.18.38807  & 17:32:12.01 & $-$29:05:16.4 &  NC & 16.951 & 18.899 & 0.21894515 & 0.685 & 0.542 & 5000.11160 \\
OGLE-BLG-ECL-000013 & BLG654.14.22809  & 17:34:24.48 & $-$29:51:38.8 &  NC & 15.535 & 17.451 & 0.21386391 & 0.314 & 0.240 & 5000.13478 \\
OGLE-BLG-ECL-000014 & BLG621.29.94695  & 17:34:59.40 & $-$21:45:07.2 &  NC & 19.471 &   --   & 0.20875910 & 0.607 & 0.491 & 5000.04820 \\
OGLE-BLG-ECL-000015 & BLG654.05.20766  & 17:35:09.37 & $-$30:12:44.8 &   C & 18.173 & 20.495 & 0.20539043 & 0.877 & 0.860 & 5000.18653 \\
OGLE-BLG-ECL-000016 & BLG672.01.10260  & 17:35:24.66 & $-$37:08:57.7 &   C & 18.716 & 21.542 & 0.21272893 & 0.497 & 0.497 & 5000.21131 \\
OGLE-BLG-ECL-000017 & BLG680.25.40021  & 17:35:43.65 & $-$37:09:58.4 &   C & 19.327 & 21.625 & 0.21383009 & 0.728 & 0.609 & 5000.11979 \\
OGLE-BLG-ECL-000018 & BLG611.11.102686 & 17:36:09.08 & $-$27:25:30.7 &  NC & 18.338 & 21.161 & 0.17470363 & 0.132 & 0.091 & 5000.08124 \\
OGLE-BLG-ECL-000019 & BLG609.25.65430  & 17:36:30.19 & $-$34:38:12.7 &  NC & 19.427 &   --   & 0.16552070 & 0.371 & 0.369 & 5000.13526 \\
OGLE-BLG-ECL-000020 & BLG653.19.19121  & 17:37:14.38 & $-$28:22:01.4 &  NC & 18.228 & 21.775 & 0.14994447 & 0.461 & 0.362 & 5000.09106 \\
OGLE-BLG-ECL-000021 & BLG680.14.44506  & 17:37:22.60 & $-$37:18:11.0 &  NC & 19.360 & 21.754 & 0.21838043 & 0.987 & 0.774 & 5000.19109 \\
OGLE-BLG-ECL-000022 & BLG680.06.29442  & 17:37:49.48 & $-$37:35:02.1 &   C & 19.106 & 21.866 & 0.21921610 & 0.721 & 0.581 & 5000.02834 \\
OGLE-BLG-ECL-000023 & BLG611.09.16725  & 17:37:53.06 & $-$27:17:40.9 &  NC & 18.285 & 21.116 & 0.20038455 & 0.285 & 0.225 & 5000.02746 \\
OGLE-BLG-ECL-000024 & BLG609.06.71216  & 17:38:08.53 & $-$35:13:25.4 &  NC & 18.657 & 20.956 & 0.21938176 & 0.304 & 0.269 & 5000.07937 \\
OGLE-BLG-ECL-000025 & BLG611.17.72634  & 17:38:11.82 & $-$27:06:05.2 &  NC & 18.787 & 22.181 & 0.17295506 & 0.233 & 0.155 & 5000.16047 \\
OGLE-BLG-ECL-000026 & BLG625.15.81252  & 17:38:14.61 & $-$22:56:53.7 &   C & 18.893 & 21.104 & 0.21928025 & 0.313 & 0.299 & 5000.16745 \\
OGLE-BLG-ECL-000027 & BLG624.23.56697  & 17:39:02.10 & $-$21:19:19.6 &  NC & 18.378 & 21.117 & 0.17040292 & 0.207 & 0.194 & 5000.04398 \\
OGLE-BLG-ECL-000028 & BLG680.12.35846  & 17:39:03.54 & $-$37:30:16.2 &  NC & 18.862 &   --   & 0.18030005 & 0.719 & 0.567 & 5000.10937 \\
OGLE-BLG-ECL-000029 & BLG675.14.69526  & 17:39:26.25 & $-$27:36:53.9 &  NC & 17.129 & 17.950 & 0.20354910 & 0.342 & 0.319 & 5000.02214 \\
OGLE-BLG-ECL-000030 & BLG625.13.72702  & 17:39:35.38 & $-$22:54:43.0 &  NC & 19.029 &   --   & 0.14745975 & 0.211 & 0.183 & 5000.06232 \\
\hline}
\end{landscape}

\begin{figure}[p!]
\includegraphics[width=12.5cm]{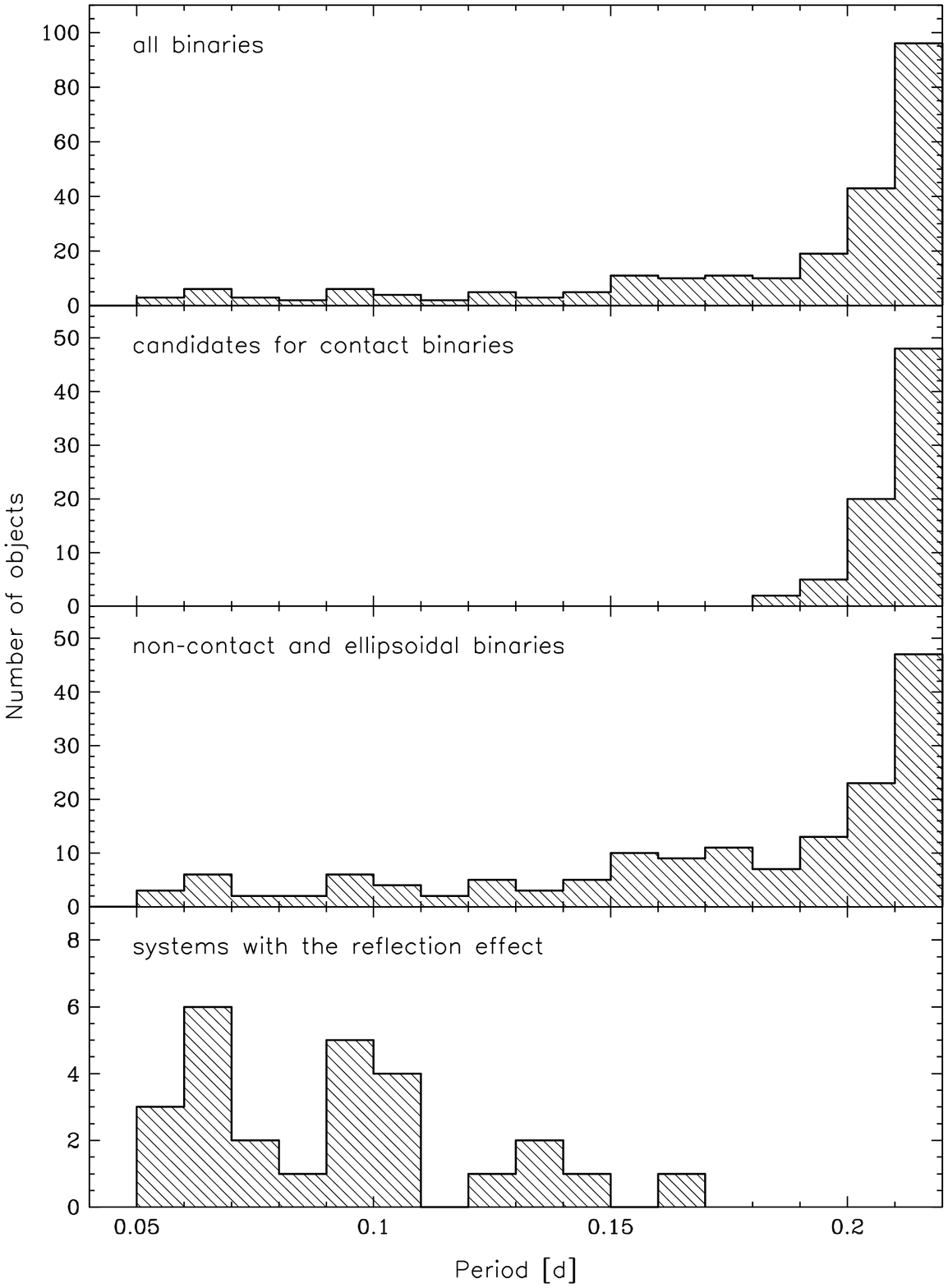}
\FigCap{Orbital period distribution of the ultra-short-period binary
systems. The consecutive panels show the period distribution of the whole
sample, the candidates for contact binary systems, non-contact binaries and
the sub-sample of eclipsing binaries with a strong reflection effect.}
\end{figure}

The vast majority of the binary systems in our sample are new
detections. Only several objects were already known from other
investigations, mostly from the previous stages of the OGLE
project. OGLE-BLG-ECL-000142 is the mentioned OGLE BW3 V38
(Udalski \etal 1995, Maceroni and Rucinski 1997). Unfortunately, in
the OGLE-IV field this star fell in a gap between two CCD chips of the
mosaic camera, so we can provide only the OGLE-III photometry of this
object. OGLE-BLG-ECL-000163 (= OGLE BUL-SC16 335) was discovered by
Po³ubek \etal (2007) and classified as an HW~Vir-type binary -- system
consisting of a B-type subdwarf (sdB) and a cool main-sequence star
(Wood \etal 1993). The spectroscopic analysis performed by Geier \etal
(2014) confirmed that the primary component is an sdB
($T_{\mathrm{eff}}=31\;500\pm1800$~K, $\log{g}=5.7\pm0.2$) and the
secondary is an M dwarf with the mass of
$0.16\pm0.05$~$M_{\odot}$. OGLE-BLG-ECL-000115 (= OGLE-I BW9 189794)
was identified by Szymañski \etal (2001) and included in their catalog
of contact binaries. OGLE-BLG-ECL-000009 and OGLE-BLG-ECL-000045 are
central stars of planetary nebulae Th 3-15 and Pe 1-9,
respectively. The latter object was identified by Mi\-szal\-ski \etal
(2009). The positions of four eclipsing binaries: OGLE-BLG-ECL-000056,
OGLE-BLG-ECL-000126, OGLE-BLG-ECL-000131, and OGLE-BLG-ECL-000201,
coincide with Novae: V825 Sco, V4742 Sgr, Nova Sgr 1986, and V5116
Sgr, respectively.

Fig.~1 shows the positions of our binary systems in the sky. Most of the
objects are likely located in front of the Galactic bulge. Note that we
detected a number of binary stars in the heavily reddened regions around
the Galactic plane ($|b|<1$). In these regions stars from the Galactic
bulge (for example RR~Lyr variables, Soszyñski \etal 2014) are
completely obscured in the visual bands by the interstellar matter.

Color--magnitude diagram for our sample is shown in Fig.~2. The apparent
$(V-I)$ colors of individual objects are dominated by the interstellar
reddening, which depends on the position in the sky and the distance to a
star, causing various types of binary systems to span a wide range of
colors. The only exclusions from this rule are binary systems exhibiting
significant reflection effect (Section~5.3), which are blue almost without
exception. In Fig.~3, we present distributions of the orbital periods for
the entire sample and separately for different types of binary systems.

\begin{figure}[p]
\includegraphics[width=12.7cm]{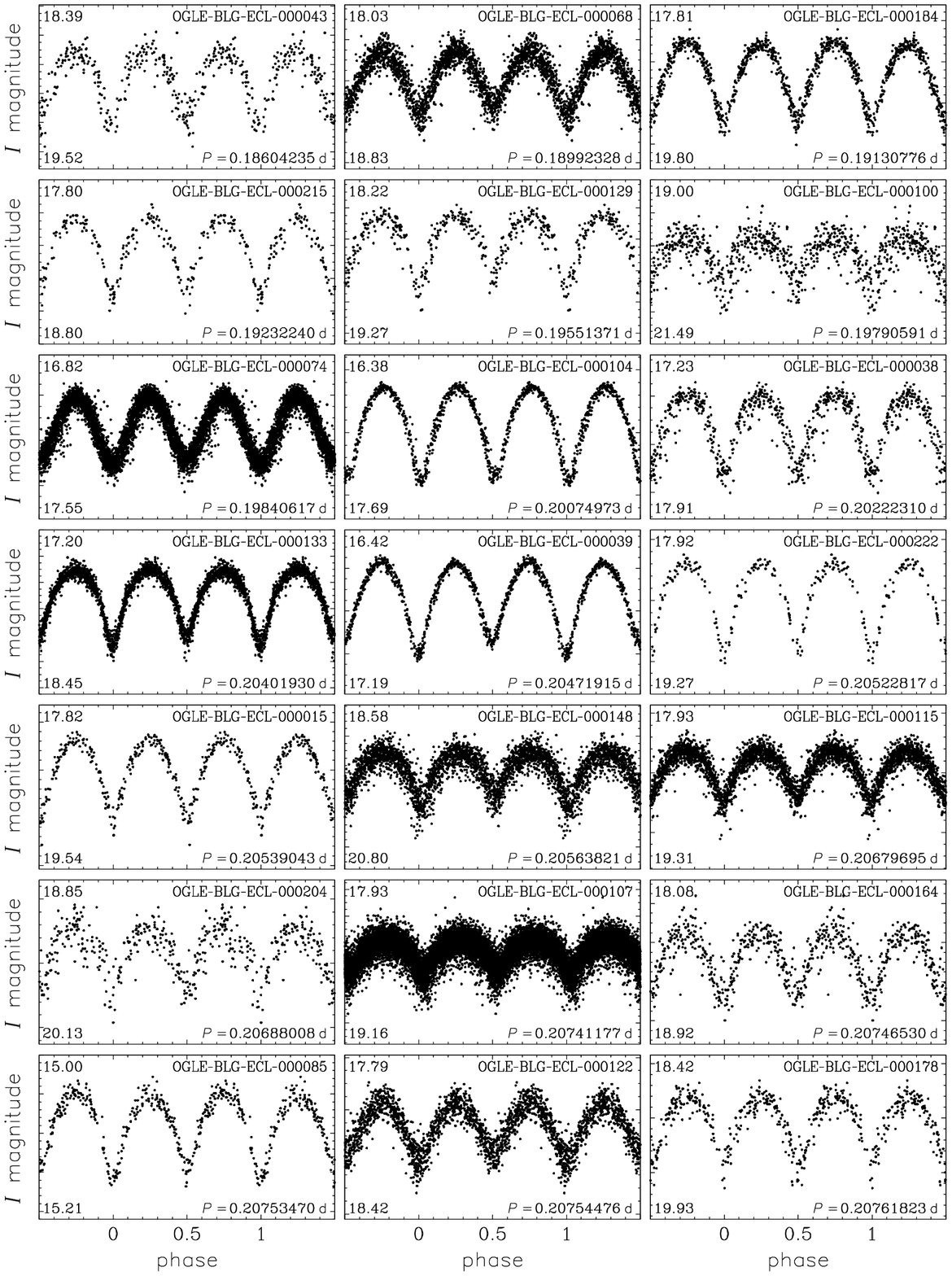}
\FigCap{{\it I}-band light curves of the shortest-period candidates for
contact binary systems. The light curves are arranged according to the
orbital periods.}
\end{figure}

\Section{Results}
\Subsection {M-dwarf Binaries and OGLE-BLG-ECL-000066}

A strong distortion of one or both components is distinctly visible in
161 light curves. 75 of them have similar depths of the primary and
secondary minima and these stars are classified as candidates for
contact binaries. The status of these objects has to be confirmed
spectroscopically. The shortest-period light curves of our candidates
for W~UMa stars are displayed in Fig.~4. The most compact binaries in
this group have $P<0.19$~d, which, if confirmed, would slightly shift
the short-period cut-off for W~UMa stars, but still this limit is much
longer than for other types of binary systems.

Our sample also contains several possible M-dwarf binaries in the detached
or semi-detached configuration with periods much below this
limit. OGLE-BLG-ECL-000066 has the period shorter than any known
main-sequence binary: 0.0984~d. Its light curve (Fig.~5) resembles those of
OGLE BW3 V38 (Maceroni and Rucinski 1997) and GSC 2314-0530 (Dimitrov and
Kjurkchieva 2010), which are M-dwarf binary systems in near-contact
configuration.

\begin{figure}[htb]
\begin{center}
\includegraphics[width=8.5cm]{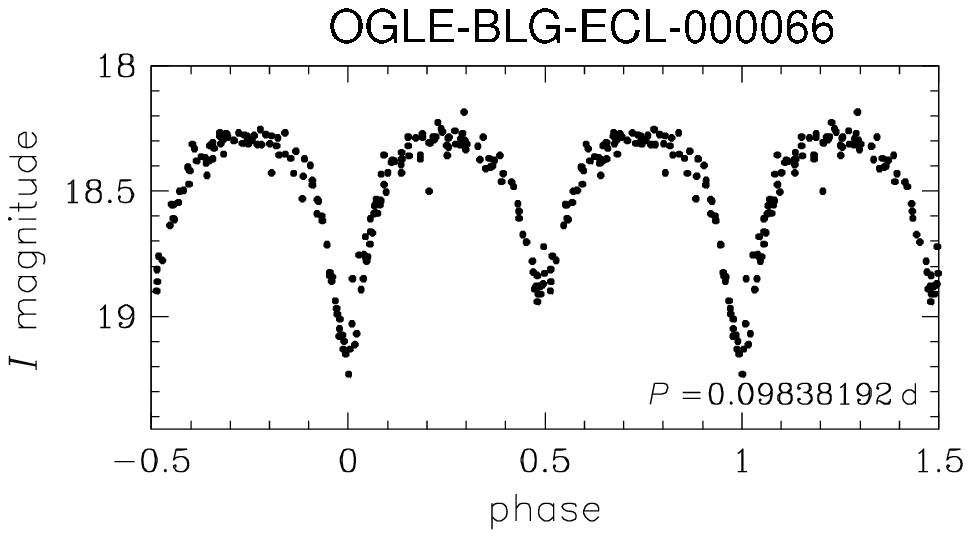}
\end{center}
\vspace*{-3mm}
\FigCap{{\it I}-band light curves of OGLE-BLG-ECL-000066 -- candidate for
the shortest-period known binary consisting of main sequence stars.}
\end{figure}

The full orbital solution for this system will be possible with the
spectroscopic measurements, although we are aware that this would be a
difficult task, since the target is faint and heavily reddened.
Moreover, its short orbital period prevents exposures longer than about
15~minutes, because of the motion blur of the spectral lines.
Nevertheless, such observations should be feasible with modern
spectrographs attached to large telescopes like X-shooter or FORS at
VLT.

However, even without spectroscopic measurements we can still
constrain the absolute parameters of the system components and attempt
to reproduce the evolutionary history of the binary. The first
constraint results from the depths of the minima. They are equal to
0.86 and 0.62~mag, suggesting that the orbit inclination $i \approx
90^{\rm o}$ and the difference in brightness of the components is of
the order of 20\%, hence the difference in mass is only about 5\%, \ie
masses are almost equal. Another constrain is connected with a low
value of the orbital period $P_{\mathrm{orb}}$. Assuming that both
components are equal mass dwarfs nearly filling their Roche lobes we
have
\begin{equation}
P_{\mathrm{orb}} > P_{\mathrm{crit}} = 0.174\sqrt{\frac{R^3}{M}}\,
\end{equation}
where $R$ and $M$ are radius and mass of each component in solar units and
$P_{\mathrm{crit}}$ is the critical period in days, when both components
just fill their Roche lobes. The observed radii of low-mass stars are
approximately equal to their masses, if both are expressed in solar units
(Torres \etal 2010). Adopting this relation we obtain $P_{\mathrm{crit}}
\approx 0.350R$ and $R < 2.857 P_{\mathrm{orb}}$. For OGLE-BLG-ECL-000066
the resulting limit for radius of each component is $R < 0.28R_{\odot}$,
equivalent to $M < 0.28M_{\odot}$. The corresponding spectral type must be
later than M3.5 (Lang 1992) but not later than M5 if the components almost
fill their Roche lobes. We adopt M4 type for both components.

These estimates are confirmed by the WD code (Wilson and Devinney
1971) fitted to the {\it I}-band light curve. Assuming two M4-type
components, the model gives masses $M_1=0.22M_{\odot}$ and
$M_2=0.21M_{\odot}$, and the mass ratio $q = 0.95$. The radii of both
components are $R_1=0.26R_{\odot}$ and $R_2=0.23R_{\odot}$ and the
inclination of the orbit $i=86^{\rm o}$. In such a configuration, the
primary component is very close to fill its Roche lobe.

The last constraint can be derived from photometry of the variable:
$V_{\mathrm{max}} = 21.5$, $I_{\mathrm{max}} = 18.285$, $J = 16.47$,
$H = 15.69$, and $K = 15.09$, all in magnitudes. The {\it V}-band
observations are at the faint-end magnitude limit of the OGLE survey
and are uncertain within $\sim0.5$~mag. The near-infrared {\it JHK}
magnitudes were obtained by the VISTA Variables in the Via Lactea
(VVV) survey (Minniti \etal 2010). The epochs of these observations
fell outside the eclipses, close to the maximum light. OGLE-BLG-ECL-000066
is located only 0.5~degrees from the Galactic plane, thus its
apparent luminosity must be severely affected by the interstellar
extinction. Indeed, its $(V-I)$ apparent color index is one of the
largest in our sample (Fig.~2).

We will estimate the amount of interstellar absorption toward our target
using its near-infrared measurements. It is known that dwarfs in the
range of spectral types from M1 to M5 have nearly the same intrinsic
$(J-H)$ color: $0.64\pm0.02$~mag (Bessell and Brett 1988). This gives
immediately the near-infrared color excess to OGLE-BLG-ECL-000066 of
$E(J-H)=0.14$~mag. It is known that the interstellar extinction toward
the Galactic bulge is anomalous (\eg Udalski 2003b, Nataf \etal 2013),
but this affects mostly the optical passbands. In the near-infrared we
can use the standard reddening law (Schlegel \etal 1998) as the first
approximation: $E(J-K)=1.62E(J-H)=0.23$~mag. This gives the dereddened
$(J-K)$ index of OGLE-BLG-ECL-000066 equal to 1.15~mag, which agrees well
with the range of this index observed for nearby M4 stars (Riaz \etal
2006).

In the optical domain, the anomalous extinction toward the Galactic bulge
makes dereddening very uncertain. The standard reddening law (Schlegel
\etal 1998) can be used to estimate an upper limit for the amount of
interstellar extinction: $A_I = 5.88E(J-H) \leq 0.82$~mag and $A_V =
10.05E(J-H) \leq 1.41$~mag. This implies that the intrinsic $(V-I)$
color of OGLE-BLG-ECL-000066 is larger than 2.6~mag, which according to
Bessell and Brett (1988) corresponds to dwarfs later than M3.5 and
earlier than M5 (assuming no extinction). This result is fully
consistent with our previous estimates.

Now we can reconstruct the evolutionary history of the binary orbit. We
assume that the only mechanism influencing the orbit evolution of
OGLE-BLG-ECL-000066 has been mass and AML due to the magnetized winds from
both components. Rates for both processes are taken from the model
described by Stêpieñ (2006a, 2011), and Gazeas and Stêpieñ (2008).

\begin{equation}
\dot M = -10^{-11}R^2\,,
\end{equation}

\begin{equation}
\frac{{\rm d}H}{{\rm d}t} = -9.8\times 10^{41}R^2M/P\,.
\end{equation}
Here $H$ is the binary angular momentum (AM) in cgs units and we assume
again equal mass components. The formulae apply to binaries in the so
called saturation regime when the stellar rotation period (equal in
this case to orbital period) is shorter than $\sim 3-4$ d (Randich \etal
1996).

\begin{figure}[t]
\includegraphics[width=12.5cm]{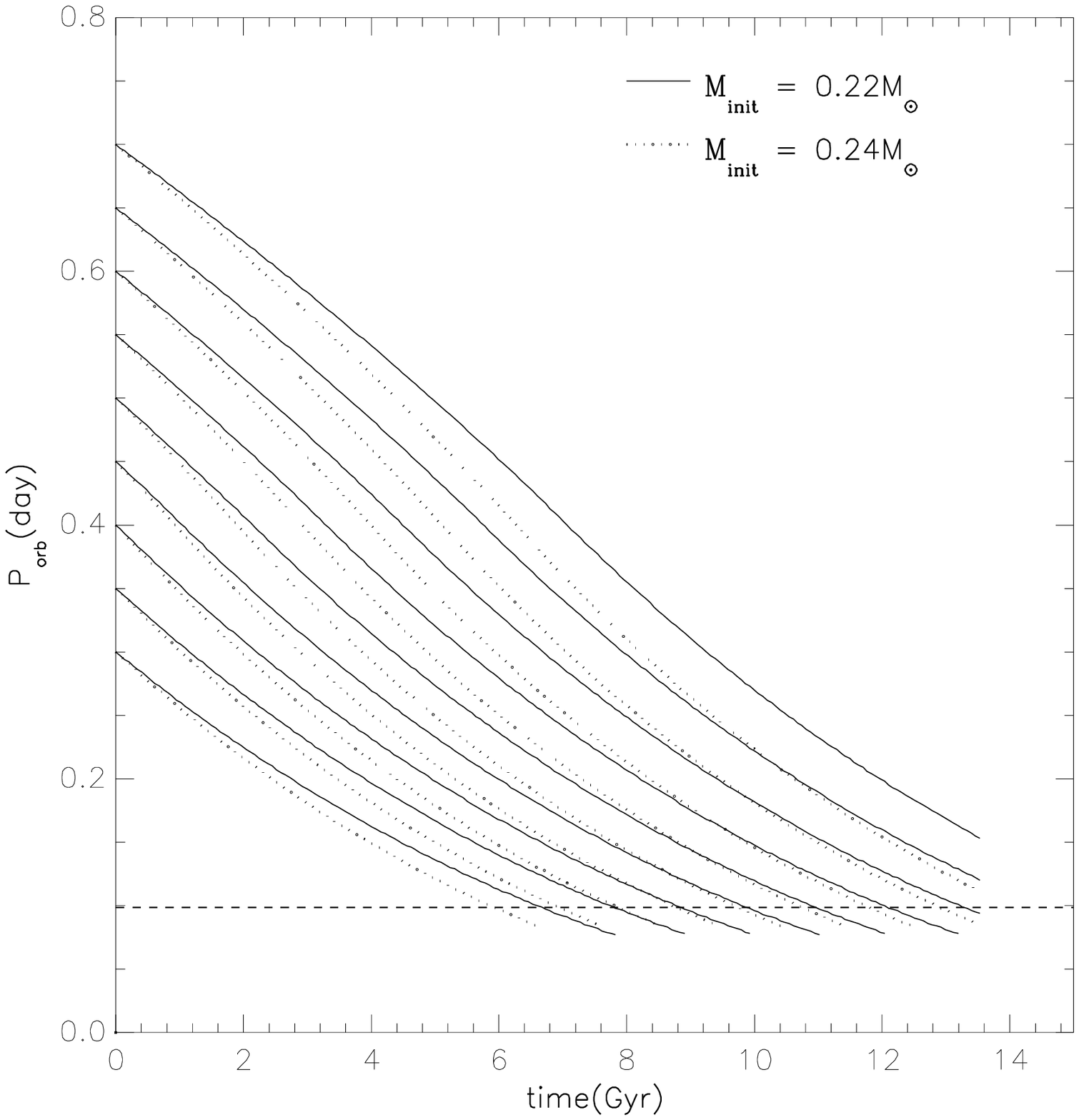}
\FigCap{Orbital period evolution of two binaries with equal mass
components and different initial periods. Solid line corresponds to 
a binary with initial component masses equal to 0.22~$M_{\odot}$ and a
dotted line to a binary with masses equal to 0.24~$M_{\odot}$. A
horizontal dashed line marks the period of OGLE-BLG-ECL-000066.}
\end{figure}

Fig.~6 presents the time evolution of the orbital period for two
binaries with initial component masses equal to 0.22$M_{\odot}$
(solid line) and 0.24$M_{\odot}$ (dotted line), and the initial
periods between 0.3-0.7~d. Due to the wind, each star loses about 2-4\%
of its original mass, before the orbital period reaches the value
observed for OGLE-BLG-ECL-000066 (dashed horizontal line). End of each
solid/dotted line marks the moment when the binary fills the Roche
lobe, or the Universe age is reached, whichever comes first. As we
see, the initial period of OGLE-BLG-ECL-000066 must have been shorter than
0.6~d, or, if its age is lower than 10~Gyr as can be judged from its
position within a thin Galactic disk, shorter than 0.4~d.

Our results demonstrated the existence of an ultra-short-period
population of near contact binaries with M-type components and periods
close to, or even below 0.1 d. AML is insufficient
during the Hubble time to reduce an initial period of 1.5 d to the
observed values. The adopted formulae are extrapolated for low-mass
stars from data on more massive stars. Can this extrapolation be
incorrect and the AML rate is in fact substantially
higher, as suggested e.g. by Jiang \etal (2012)? We calculated several
evolutionary models of binaries with initial component masses equal to
0.24~$M_{\odot}$ and AML rate increased arbitrarily by a factor of 5
and 10. Binaries with the initial period of 1.5~d reach the value
observed for OGLE-BLG-ECL-000066 in 6.6 and 3.26~Gyr, respectively. Even
binaries with the initial period of 2.5~d reach this value within the
Hubble time. However, the AML rate of single M stars does not seem to
be so high. In fact, the analysis of rotation of M-type stars by
Delfosse \etal (1998) indicates that the time scale for spin-down of
single stars increases with the decreasing mass from 50~Myr for G
dwarfs to a few Gyr for M3-M4 stars. The same ratio of both time
scales can be obtained from Eq.~3. Barnes and Kim (2010)
derived an empirical expression for the AML rate of cool stars with
different masses. The rate is proportional to the ratio of stellar
moment of inertia and a parameter called turnover time. This ratio
decreases by a factor of $\sim 10^2$ between 1 and 0.2 solar mass star
with no sign of increase for low mass stars. So, the substantially
increased AML rate of M-type stars during the main sequence
evolutionary phase finds no substantiation.

Another possibility is that M-stars lose a substantial fraction of
their AM during the T~Tau phase. The life-time of a cool star in
this phase is roughly equal to 1\% of the main sequence life time. For
solar type stars it amounts to only about $10^7$--$10^8$~years -- short
enough to neglect the shortening of the orbital period even if AML
rate in T~Tau phase for a single star is significantly higher than
in the main sequence phase. This may not be true, however, for
binaries with component masses 0.2--0.3~$M_{\odot}$ which spend
$\sim1$~Gyr before they reach the zero age main sequence. Little is known
about activity of low mass T~Tau stars and their interaction with
circumstellar (or circumbinary) disks so we cannot exclude this
possibility. Observations of young M-type binaries would help to
follow the period evolution in the early evolutionary phases.

Alternatively, we can consider dynamical interaction with other stars.
Naoz and Fabrycky (2014) computed orbit evolution of a numerous sample
of binaries under the influence of the KCTF mechanism, including
relativistic effects and so called octupole-level approximation. The
resulting period distribution of the inner binaries contains a
fraction of periods shorter than 1 d. A majority of them originate
from binaries with equally short initial periods because the authors
adopted the initial period distribution after Duquennoy and Mayor
(1991), which extends smoothly to zero. By adopting a cut-off period of
1.5-2~d we should reject these binaries as nonphysical but a very limited
number of longer period binaries also evolved to, or below 1~d limit
(see their Fig.~3). We conclude that the KCTF mechanism can
produce binaries with periods of 1~d or less under very exceptional
circumstances. Based on simple statistics, Drake \etal (2014) argue
that dM+dM ulta-short-period binaries are indeed extremely rare among
photometrically variable stars. To verify this hypothesis the
bias-free period distribution of cool binaries should be obtained.

\Subsection{Period changes}

The detection of period changes in close binary systems would be a
powerful test for models explaining the short-period cut-off in
contact binaries. Generally, the orbital periods should decrease over
time, as the systems lose angular momentum from magnetic braking,
which ultimately leads to the coalescence. So far the only stellar
merger observed before and during the act of coalescence was V1309 Sco
(Tylenda \etal 2011), photometrically monitored by the OGLE survey.

Kubiak \etal (2006) detected statistically significant secular period
changes in 134 of 669 analyzed contact binaries observed by OGLE toward the
Galactic bulge (Szymañski \etal 2001). The distribution of the rates of the
period changes occurred to be nearly symmetrical around zero. Pilecki \etal
(2007) analyzed 1711 contact and semi-detached binaries observed by the ASAS
project and found 31 period-changing systems -- 21 decreasing and 10
increasing periods. Lohr \etal (2013) found statistically significant
($3\sigma$) period changes in 38 of 143 short-period binaries observed by
the SuperWASP project. The distribution of the period-change rates was
approximately symmetrical around zero.

We used the fitted model light curves to perform the analysis of the
orbital period changes for all eclipsing binaries in our sample,
although reliable results can be obtained only for the stars that were
observed during both -- OGLE-III and OGLE-IV -- stages of the
survey. Other light curves have too short time span to draw
statistically significant conclusions from the $(O-C)$ diagrams. Our
procedure was as follows: each light curve was divided into separate
observing seasons (spanning one year of observations), and for each
such chunk we shifted the phase of the model light curve to minimize
the sum of squared differences between the observed and the model
light curve. In such a way we obtained $(O-C)$ diagrams for our stars
with one point per every observing season.

\begin{figure}[t]
\includegraphics[width=12.7cm]{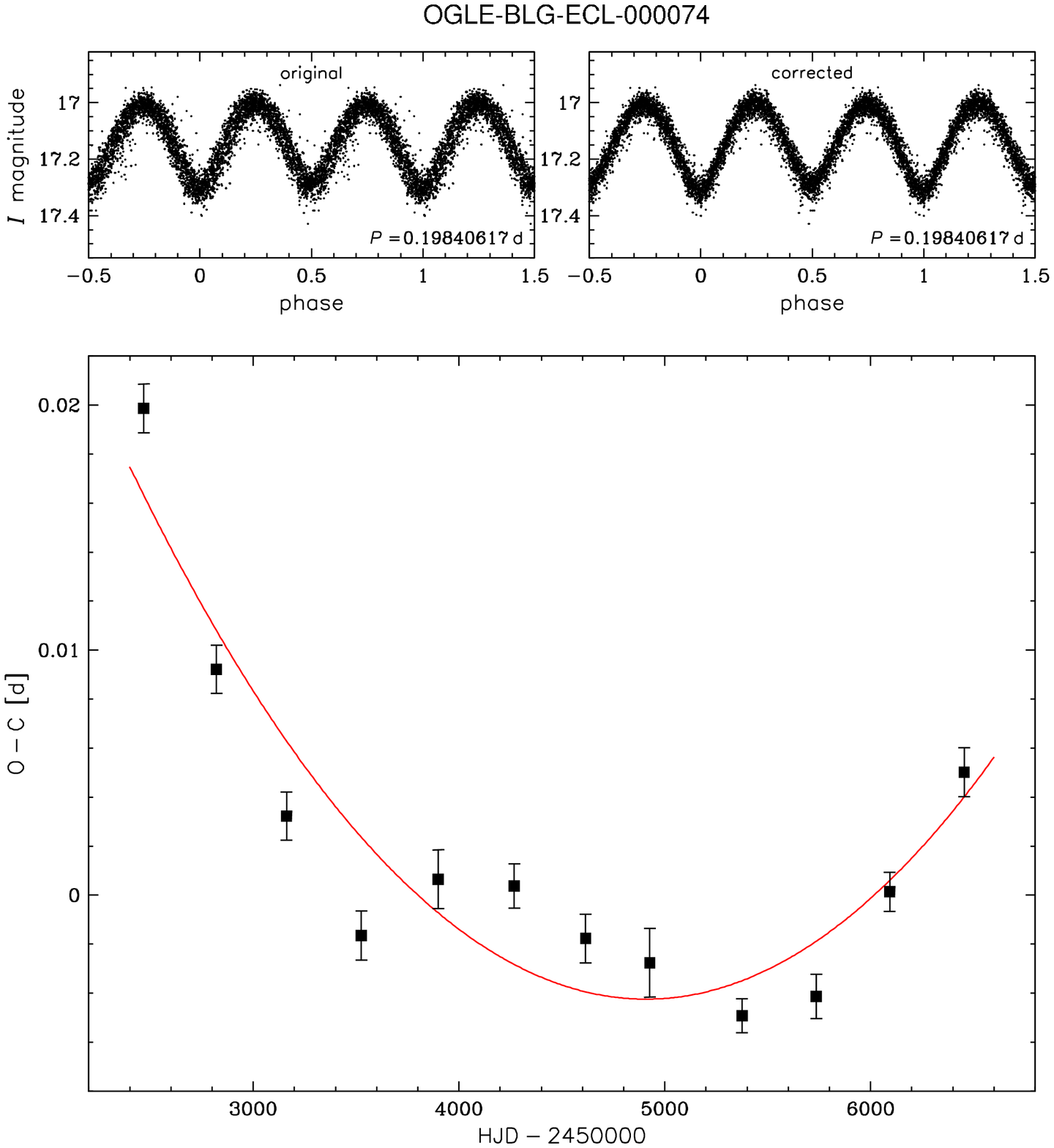}
\FigCap{{\it Upper panels}: light curve of a compact binary system
OGLE-BLG-ECL-000074. {\it Left panel} presents the original data
folded with the mean orbital period, while {\it right panel} presents
the light curve corrected for the monotonic period increase measured
in the $(O-C)$ diagram ({\it lower panel}).}
\end{figure}

\renewcommand{\TableFont}{\scriptsize}
\MakeTable{l@{\hspace{8pt}}
c@{\hspace{6pt}}
c@{\hspace{3pt}}
r@{\hspace{10pt}}}
{12.5cm}{Eclipsing binary systems with monotonic period changes.}
{\hline
\multicolumn{1}{c}{Identifier} 
&\multicolumn{1}{c}{Type} 
&\multicolumn{1}{c}{$P_{\mathrm{orb}}$} 
&\multicolumn{1}{c}{$\mathrm{d}P_{\mathrm{orb}}/\mathrm{d}t$} \\
& & \multicolumn{1}{c}{[d]}
&\multicolumn{1}{c}{[s/year]} \\
\hline
OGLE-BLG-ECL-000065 &  C & 0.21221016 & $-$0.04 \\
OGLE-BLG-ECL-000073 &  C & 0.21831428 &    0.17 \\
OGLE-BLG-ECL-000074 &  C & 0.19840617 &    0.22 \\
OGLE-BLG-ECL-000131 & NC & 0.15356160 &    0.29 \\
OGLE-BLG-ECL-000141 &  C & 0.21436650 & $-$0.13 \\
OGLE-BLG-ECL-000144 &  C & 0.21894593 &    0.17 \\
OGLE-BLG-ECL-000145 &  C & 0.21104030 &    0.10 \\
OGLE-BLG-ECL-000148 &  C & 0.20563821 &    0.14 \\
OGLE-BLG-ECL-000168 & NC & 0.20280139 & $-$0.15 \\
OGLE-BLG-ECL-000173 & NC & 0.21373651 &    0.38 \\
\hline}

Among 78 binary systems observed by OGLE for over 12 or more years, we
found 16 objects with apparently unstable periods. In 10
cases, the changes of periods are roughly monotonic, like in the case
of OGLE-BLG-ECL-000074 shown in Fig.~7. Table~2 lists the binary systems
with secular period changes. Seven systems increase and three systems
decrease their orbital periods. Most of these objects are candidates
for contact or nearly contact binary systems. The shortest-period
binary changing its period is OGLE-BLG-ECL-000131 = Nova Sgr 1986. In six
systems listed in Table~3, the orbital period changes are likely not
constant. Since the typical time scales of the period variations are
of the order of several years, in most cases we cannot unambiguously
distinguish whether these variations are strictly periodic or they are
irregular in nature. Cyclic period variations may reveal the presence
of an unseen tertiary companion due to the light travel time
effect. Irregular period fluctuations may be caused by non-stationary
mass transfer within the system or by mass ejections from the
system. Table~3 gives possible lengths of the period change cycles.

\renewcommand{\TableFont}{\scriptsize}
\MakeTable{l@{\hspace{8pt}}
c@{\hspace{6pt}}
c@{\hspace{3pt}}
c@{\hspace{1pt}}}
{12.5cm}{Eclipsing binary systems with cyclic or irregular period changes.}
{\hline
\multicolumn{1}{c}{Identifier} 
&\multicolumn{1}{c}{Type} 
&\multicolumn{1}{c}{$P_{\mathrm{orb}}$} 
&\multicolumn{1}{c}{$P_{\mathrm{cycle}}$} \\
& & \multicolumn{1}{c}{[d]}
&\multicolumn{1}{c}{[d]} \\
\hline
OGLE-BLG-ECL-000069 &  NC & 0.19272500 &  2300 \\
OGLE-BLG-ECL-000104 &   C & 0.20074973 &  1400 \\
OGLE-BLG-ECL-000107 &   C & 0.20741177 &  2400 \\
OGLE-BLG-ECL-000127 &  NC & 0.16641697 &  1500 \\
OGLE-BLG-ECL-000170 &   C & 0.21491916 &  3300 \\
OGLE-BLG-ELL-000015 & ELL & 0.21996525 &  2200 \\
\hline}

\begin{figure}[p]
\includegraphics[width=12.7cm]{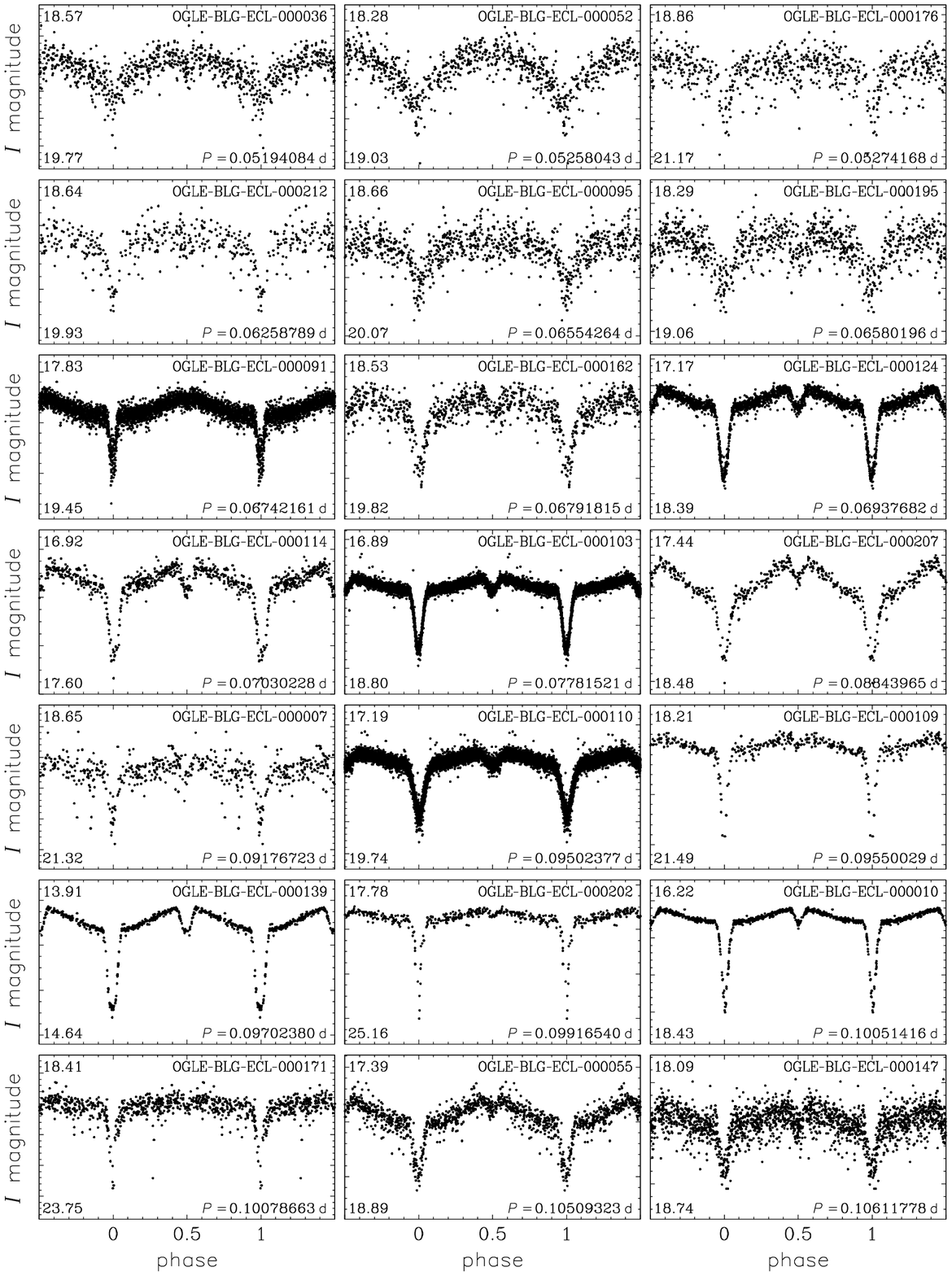}
\FigCap{{\it I}-band light curves of the shortest-period eclipsing
binary systems with a strong reflection effect. The light curves are
arranged according to the increasing orbital periods.}
\end{figure}

\Subsection{Eclipsing Binaries with a Strong Reflection Effect}
Among detached eclipsing binary systems, 26 objects constitute a
homogeneous group. The light curves of 21 shortest-period objects from
this group are shown in Fig.~8. These stars are characterized by narrow
eclipses with very different depths of the primary and secondary eclipses
and by the sinusoidal modulation between the eclipses caused by the
reflection of light received by the cooler star from the hotter
component. Ten similar systems were recently reported in the OGLE-III
Galactic disk fields (Pietrukowicz \etal 2013), six of which have orbital
periods shorter than 0.15~d. Objects presented in Fig.~8 belong to the
bluest and the shortest-period binary systems in our collection (see Fig.~2
and the lower panel of Fig.~3). These systems consist likely of a cool main
sequence star and a hot evolved remnant: an sdB or a white dwarf (\eg
{\.I}bano{\v g}lu \etal 2004, Zorotovic and Schreiber 2013). Their short
periods suggest that they are post common-envelope systems (Paczyñski
1976), so they are important for understanding common envelope evolution
and subsequent system behaviors. Due to their well-defined, deep primary
eclipses and short orbital periods, such objects have also great potential
in detecting circumbinary substellar companions. At least a dozen compact
binaries containing a white dwarf or an sdB primary have been claimed to
host a planetary or brown dwarf third companion from the eclipse timing
measurements (\eg Guinan and Ribas 2001, Lee \etal 2009, Beuermann \etal
2010), however the reality of some of these detections is controversial
(Lohr \etal 2014). Our preliminary timing analysis showed no plausible
evidence for period changes in any binary system with a strong reflection
effect.

\Subsection{Cataclysmic Variables}

Cataclysmic variables (CVs) are interacting binary systems consisting
of a white dwarf and a mass transferring secondary (donor), usually a
low-mass main-sequence star. If the magnetic field of the white dwarf
is weak, the gas flowing through the inner Lagrangian point forms an
accretion disk around the primary. The location where the material
hits the edge of a disk is called the hot spot. In dwarf novae,
instabilities in the disk lead to regular outbursts with typical
amplitudes of 2--5 mag.

\begin{figure}[t]
\includegraphics[width=12.7cm]{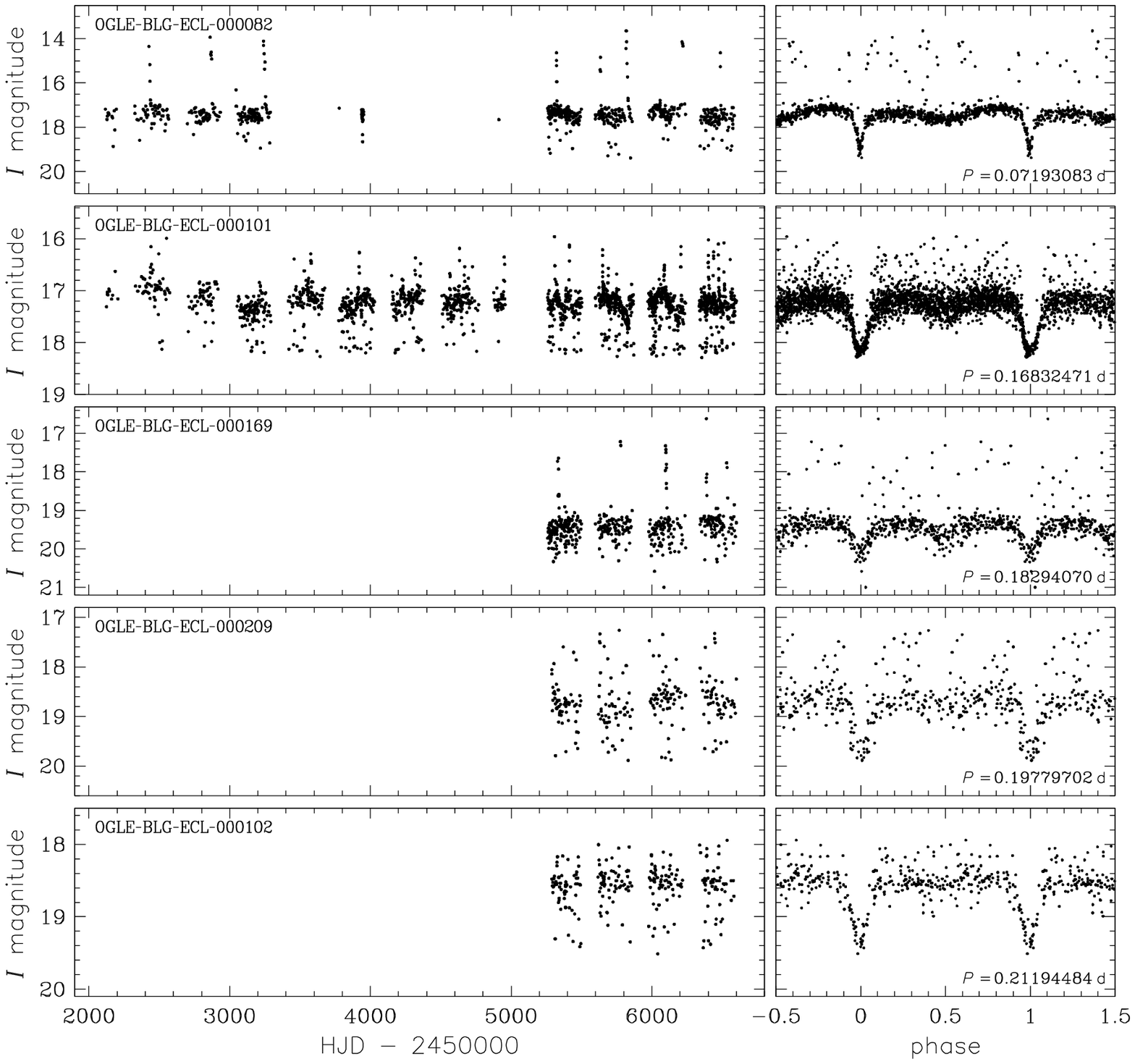}
\FigCap{{\it I}-band light curves of dwarf novae in our sample. {\it Left
panels} present unfolded light curves, {\it right panels} show the same
light curves folded with the orbital periods.}
\end{figure}

About 20\% of known CVs are eclipsing binaries (Warner 1995). An
analysis of a complex eclipse shape may provide information about the
relative brightness, sizes and masses of both components and the hot
spot (\eg Wood \etal 1986, Horne \etal 1994). The distribution of
orbital periods reflects the evolution of CVs.

\begin{figure}[t]
\includegraphics[width=12.7cm]{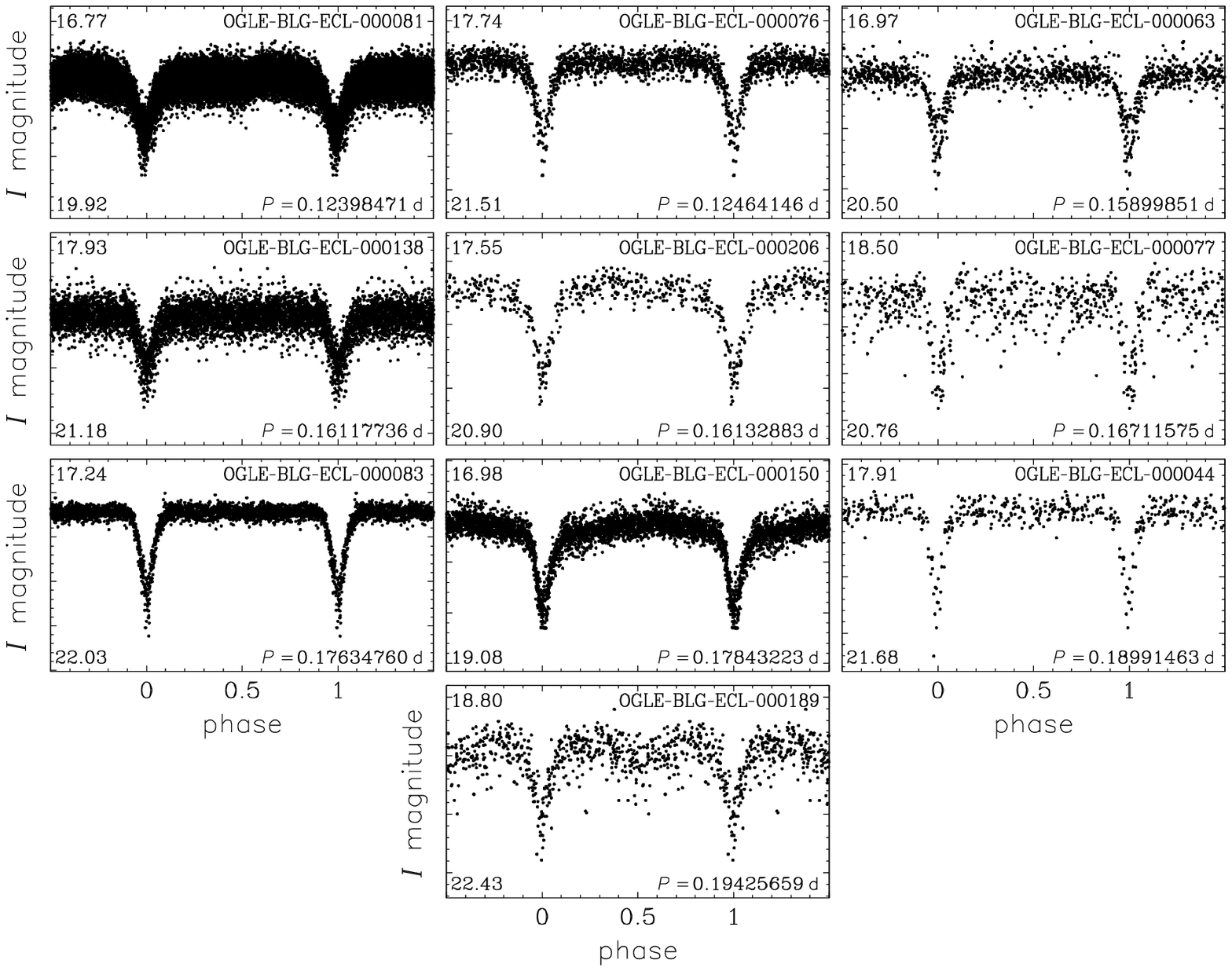}
\FigCap{{\it I}-band light curves of candidates for cataclysmic variables.}
\end{figure}

Our sample contains five eclipsing dwarf novae. Their unfolded and
folded light curves are showed in Fig.~9. OGLE-BLG-ECL-000082 (with the
orbital period of 1.726~h) probably belongs to SU~UMa-type dwarf novae
with brighter and longer superoutbursts every $\sim 420$~d. There are
only a handful of eclipsing SU~UMa stars known. Analysis of their
light curves may give evidence for the origin of superoutbursts (\eg
Smak 1994, B\k{a}kowska and Olech 2014). The remaining four objects
have orbital periods longer than 4~h. They are likely members of the
U~Gem class. The outbursts have various amplitudes and recurrence
intervals, while the eclipses are always deep, exceeding 0.85 mag in
the $I$-band, but with no clear reflection effect.

Ten further detached eclipsing systems exhibit similar depths of
minima (in some cases exceeding 2~mag), but with no outbursts
(Fig.~10). Secondary eclipses are very shallow, sometimes invisible,
and distinct reflection effect is absent. Some of these objects show
irregular variations outside the eclipses (flickering), produced by
the inner disk and/or the hot spot. These are probably UX~UMa-type
stars, \ie Nova and Nova-like cataclysmic variables (\eg Smak
1994). Spectroscopic observations would provide conclusive evidence
about the nature of the remaining objects.

\begin{figure}[t]
\begin{center}
\includegraphics[width=8.5cm]{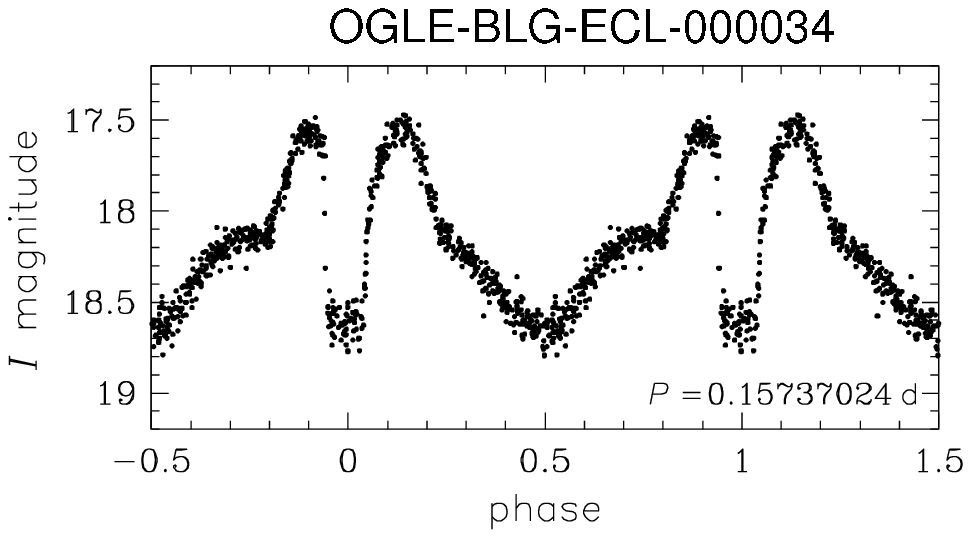}
\end{center}
\vspace*{-3mm}
\FigCap{{\it I}-band light curve of OGLE-BLG-ECL-000034 -- a candidate
for a high inclination polar.}
\end{figure}

\Subsection{The Case of OGLE-BLG-ECL-000034}

The most bizarre light curve in our collection is shown by
OGLE-BLG-ECL-000034 (Fig.~11). Deep eclipses ($\Delta I = 1.1$ mag)
recur every 3.777~h. The out-of-eclipse brightness is variable. Between 
phases 0.25 and 0.8 the light curve has sinusoidal shape, while
between 0.8 and 0.9 (0.15 and 0.25) we observe steep brightening
(fading). The light curve is almost symmetric. The ephemeris for the
mid-eclipse are:
\begin{displaymath}
{\rm HJD_{eclipse}}=2\;455\;000.082(1)+0.15737024(4)\cdot E.
\end{displaymath}

The observed changes could be explained, if OGLE-BLG-ECL-000034 were a
high inclination polar. Polars (AM~Her-type stars) are cataclysmic
variables in which the magnetic field of the white dwarf is very
strong ($B>10^3$T at the surface). The accretion disk cannot be formed
and the material from the secondary follows magnetic field lines,
heading toward magnetic pole(s). Accretion regions around magnetic
pole(s) are main sources of the radiation from the system. The strong
magnetic field also synchronizes the rotation of the white dwarf and
the binary.

To account for the light curve properties, the inclination of the
system has to be very high and the magnetic axis of the white dwarf
must point toward the secondary. In such a case, the gas flows
preferentially onto the closer magnetic pole. This pole is hidden
behind the white dwarf between phases 0.25 and 0.8. Bumps between 0.8
and 0.25 can be naturally explained by the emergence of the accretion
region. It is a source of cyclotron radiation; possibly, a cyclotron
hump peaks around 800~nm, giving rise to the large amplitude in the
$I$-band. The light curve of the well-known polar, EP~Dra, shows
similar features (see Remillard \etal 1991 and their Fig.~6). The red
dwarf secondary, filling its Roche lobe, is strongly distorted,
producing an additional sinusoidal signal at half of the orbital
period.

We are able to assess basic system parameters from the light curve
examination. Assuming that the secondary fills its Roche lobe and it
follows the standard mass-radius relation for red dwarfs, a mass of
the secondary is $M_2\approx 0.065 P^{5/4}_{\rm orb}\ M_{\odot} =
0.34\ M_{\odot}$ (Hellier 2001). The eclipse width is
$\Delta\varphi\approx 0.1$, suggesting a relatively high mass ratio
$q\gtrsim 0.5$ (Horne 1985). For the inclination $i=85^{\circ}$, the
white dwarf mass is $M_1\approx0.6\ M_{\odot}$ and the binary
separation $a\approx 1.2\ R_{\odot}$. Lower inclination,
$i=80^{\circ}$, corresponds to the less massive white dwarf
$M_1\approx 0.4\ M_{\odot}$ and slightly smaller separation $a\approx
1.1\ R_{\odot}$.

We checked that there is no X-ray counterpart to this source in the
HEASARC\footnote{http://heasarc.gsfc.nasa.gov} database. Further
spectroscopic and polarimetric observations are needed to testify our
hypothesis.

\Section{Conclusions}
We presented a sample of 242 ultra-short-period binary systems
detected toward the Galactic bulge by the OGLE survey. Our collection
significantly increases the number of known binary systems with
periods below 0.22~d. Together with the list of objects we provide
their long-term time-series photometry in two filters. The sample is
very heterogeneous -- it contains candidates for contact binaries,
semi-detached and detached systems, dwarf novae, and HW~Vir stars. One
of our objects -- OGLE-BLG-ECL-000066 -- is a candidate for the
shortest-period known binary with non-degenerate components. The
existence of such systems is a challenge for the binary evolution
theory.

The presented sample is a forerunner of the OGLE collection of
eclipsing and ellipsoidal binary systems toward the Galactic bulge
which will be published in the future. Binary stars with the orbital
periods above the adopted limit -- 0.22~d -- are among the most
numerous variable stars. We expect that the total number of eclipsing
stars in the OGLE collection should exceed one hundred thousand.

\Acknow{We are grateful to Z.~Ko³aczkowski and A.~Schwarzenberg-Czerny
for providing software which enabled us to prepare this study.

The OGLE project has received funding from the European Research
Council under the European Community's Seventh Framework Programme
(FP7/2007-2013)/ERC grant agreement no. 246678 to AU. This work has
been supported by the Polish National Science Centre grant
No. DEC-2011/03/B/ST9/02573. We gratefully acknowledge financial
support from the Polish Ministry of Science and Higher Education
through the program ``Ideas Plus'' award No. IdP2012 000162. KS
acknowledges the financial support of the National Science Center
through the grant DEC-2011/03/B/ST9/03299.}

\end{document}